# Direct observation of terahertz frequency comb generation in difference-frequency quantum cascade lasers

Luigi Consolino[1,*], Malik Nafa[1], Michele De Regis[1], Francesco Cappelli[1], Saverio Bartalini[1], Akio Ito[2], Masahiro Hitaka[2], Tatsuo Dougakiuchi[2], Tadataka Edamura[2], Paolo De Natale[1], Kazuue Fujita[2,*]

[1]CNR-Istituto Nazionale di Ottica and LENS, Via N. Carrara 1, 50019 Sesto Fiorentino (FI), Italy.

[2]Central Research Laboratory, Hamamatsu Photonics K.K., 5000 Hirakuchi, Hamakita-ku, Hamamatsu 434-8601, Japan.

[*]luigi.consolino@ino.cnr.it, kfujita@crl.hpk.co.jp

**Terahertz quantum cascade laser sources based on intra-cavity difference frequency generation from mid-IR devices are an important asset for applications in rotational molecular spectroscopy and sensing, being the only electrically pumped device able to operate in the 0.6-6 THz range without the need of bulky and expensive liquid helium cooling. Here we present comb operation obtained by intra-cavity mixing of a distributed feedback laser at $\lambda = 6.5$ µm and a Fabry-Pérot device at around $\lambda = 6.9$ µm. The resulting ultra-broadband THz emission extends from 1.8 to 3.3 THz, with a total output power of 8 µW at 78K. The THz emission has been characterized by multi-heterodyne detection with a primary frequency standard referenced THz comb, obtained by optical rectification of near infrared pulses. The down-converted beatnotes, simultaneously acquired, confirm an equally spaced THz emission down to 1 MHz accuracy. In the next future this setup can be used for Fourier transform based evaluation of the phase relation among the emitted THz modes, paving the way to room-temperature, compact and field-deployable metrological grade THz frequency combs.**

## Introduction

In the field of frequency metrology, a paradigmatic shift has occurred with the invention of optical Frequency Combs [1–3] (optical FCs), whose emission is a superposition of a series of quasi-monochromatic electromagnetic waves (the FC modes) that can be fully controlled via two parameters, i.e. repetition rate and offset frequency. To this purpose, a necessary and non-trivial requirement is the existence of a fixed (i.e. constant over time) phase relation among all the modes, defining the overall coherence of the source. In this way, the modes of such lasers can be used as a precise ruler in the visible and near-infrared (IR) domain, providing a direct link between optical and microwave/radio frequencies [4]. Thanks to these outstanding features, FCs are nowadays key tools in many fields of fundamental and applied research [5]. Moreover, in recent years FCs' application possibilities have been migrated to less common spectral regions (mid- and far-IR, ultraviolet), while vigorous efforts have been spent trying to miniaturize these sources, achieving the most interesting results with mainly three different technologies: microresonators [6,7], interband cascade lasers [8] and quantum cascade lasers frequency combs (QCL-FCs) [9]. However, among these, only QCL-FCs are able to access the THz portion of the electromagnetic spectrum [10,11].

This THz (or far-infrared) window, ranging from 0.1 to 10 THz, is a largely underexploited section of the electromagnetic spectrum, and it has been historically referred to as "THz gap". However, in the latest years high-precision molecular spectroscopy in this region has attracted a lot of attention from the scientific community, as many intriguing molecules present characteristic rotational and ro-vibrational THz transitions that, as a consequence, build up a novel fingerprint region [12]. For this reason, in order to be suitable for metrological-grade THz spectroscopic applications, a quest for high performance laser sources has been initiated [13]. These lasers sources should be ideally spectrally pure, broadband or widely tunable, traceable against a primary frequency standard, emitting high-power, and - last but not least - compact and easy to operate for *in-situ* or field operation.

In this respect, single frequency THz QCLs might be ideal candidates. In fact, these current-driven semiconductor lasers rely on intersubband transitions in quantum wells, and therefore their emission frequency can be quantum engineered. Moreover, they show relatively high optical power [14–19] and very high spectral purity [20–22]. These features have been exploited for high precision THz molecular spectroscopy [23,24] by phase referencing a QCL to a THz FC [25,26]. Unfortunately, this approach still shows two main limitations: firstly, the limited mode-hop-free tunability range of the single device (few hundred MHz) is not suited for broadband spectroscopy; secondly, the need of cryogenic cooling has profoundly hampered QCLs' widespread use. In fact, despite the extreme miniaturization of the sources, the need of a liquid helium cryostat, of the corresponding liquid helium dewar, or (alternatively) the need of expensive low mechanical noise pulse tube refrigerators, make the experimental setup bulky, unfriendly to operate, and unfit for field deployment, even if first attempts have been recently pioneered [17,27,28].

The first problem can be efficiently solved by replacing QCL single frequency devices with QCL-FCs based on broadband Fabry-Pérot (FP) devices with low group velocity dispersion [9,10]. In fact, thanks to degenerate and non-degenerate four wave mixing non-linear processes inside the device active region, the cavity longitudinal modes can be injection-locked, resulting in a coherent FC emission. In the THz range, two different techniques have been adopted to demonstrate the high level of coherence achieved by QCL-FC sources: shifted wave interference Fourier transform spectroscopy (SWIFTS) [29] and, more recently, Fourier analysis of comb emission (FACE) [30,31]. Thanks to these properties, QCL-FCs can be employed in spectroscopic setups, such as dual-comb spectroscopy (DCS) setups, as proposed in 2016 with an etalon signal [32] (i.e. simulating a real molecular absorption), with spectra of ammonia gas [33] and water vapour [34,35], and even exploiting frequency referencing in a hybrid DCS spectrometer [36].

The second huge limitation of THz QCLs is the need of cryogenic cooling, ultimately hindering the miniaturization of QCL-based setups, which can now be overcome with an alternative approach based on difference frequency generation (DFG) in mid-infrared QCLs [37], referred as THz DFG-QCLs [38–40]. Mid-infrared QCLs are engineered to provide mid-infrared gain for pumps and to possess giant second-order nonlinearity $\chi^{(2)}$ for THz DFG in the QCL active region [38–41]. Since nonlinear processes, such as DFG, do not require any population inversion, THz DFG-QCLs are able to operate at room temperature, similar to other mid-infrared QCLs. As a consequence, these are currently the only electrically pumped, monolithic semiconductor THz sources operable at room temperature in the 0.6 to 6 THz frequency range [42,43]. This key operation mode has strongly motivated, in the last years, further research on DFG-based QCL lasers, although the achievable emitted power is lower than in directly THz emitting lasers. By adopting a strongly coupled upper-state (dual-upper-state: DAU) active region design approach [41,44], which does not need two stacked laser active regions for dual wavelength mid-IR pumps, continuous-wave (CW) performance of THz DFG-QCLs has considerably improved in the past few years [45,46].

In the initial research and development of THz DFG-QCLs, narrow-linewidth, widely tunable devices have been extensively developed. The potential use of single mode devices as a metrological grade reference for heterodyne detection has also been investigated [47]. This technology has been recently migrated to DFG QCL devices operating in multimode regime, in which broadband THz emission is generated via nonlinear mixing between a single mid-IR pump frequency selected by a largely detuned distributed feedback (DFB) grating, and FP modes of the second mid-IR pump selected by the laser cavity [48]. The potential operation of THz combs has been assessed for multimode THz DFG-QCL devices, initially at 78 K [40], and subsequently at room temperature [49]. However, these were evaluated against the spectral coherence of mid-IR emission spectra, and by retrieval of a single and narrow intermodal beatnote (IBN). However, this is a condition that is necessary, but not sufficient to prove comb operation, as the difference between multimode operation and comb emission ultimately relies in the equal mode spacing, and a phase relation (binding the modes one another) being constant over time.

In this work, we demonstrate the detection of the modes emitted by a DFG QCL-FC with a multi-heterodyne technique, which eventually returns a resolution much higher than Fourier transform based spectrometers (in the order of few GHz). This enables a direct comparison of the mode spacing with the electrically detected IBN frequency, which results in a very good agreement. Unfortunately, the signal to noise ratio provided by the experimental setup, limited by the THz power-per-mode emitted by the DFG QCL device does not allow for a characterization of the device phase relation, that will be attempted with next generation devices.

## THz DFG-QCL design and characterization

The DAU active region structure is used for the present DFG-QCL, in which the intersubband transitions for laser action occur from two anti-crossed upper subbands to lower miniband and these many transitions result in a broad-gain spectrum. In the active region designs, almost equal dipole matrix elements are designed for the inetersubband transitions from the two upper states to lower laser states in order to achieve the flat-top broad electroluminescence emission spectrum. In the DAU active region, several sets of states contribute to resonant $\chi^{(2)}$ for DFG. These energy states relevant to the optical resonance were engineered to attain the second order optical nonlinearity in the DAU active region. For the active region of a THz-DFG device emitting around 3 THz, the estimated module of the nonlinear susceptibility is $|\chi^{(2)}| = 7.8$ nm/V.

The waveguide core in our CW devices consists of 40-staege active regions (sheet doping density in each active region stage: $1.0 \times 10^{11}$ cm$^{-2}$). The growth of all the semiconductor layer structures were done by metal organic vapor phase epitaxy method on an undoped InP substrate [50,51]. The waveguide structure was designed to achieve optical mode confinement for mid-infrared, and THz DFG emission at a Cherenkov phase matching angle of ~ 20 degrees into the undoped InP device substrate. A schematic of the device structure is shown in Fig. 1(a). The growth initiates with a 200 nm thick $In_{0.53}Ga_{0.47}As$ current injection layer (Si, $1.0 \times 10^{18}$ cm$^{-3}$) and then a 5 μm thick n-InP (Si, $1.5 \times 10^{16}$ cm$^{-3}$) is formed as a lower cladding layer. The strain compensated InGaAs/InAlAs active region layers are sandwiched between n-$In_{0.53}Ga_{0.47}As$ guide layers (Si, $1.5 \times 10^{16}$ cm$^{-3}$) where the thicknesses of 250 nm and 450 nm are used for lower and upper layers. A buried DFB grating (single-period) was defined by nanoimprint lithography for the single-mode laser emission and etched into upper n-$In_{0.53}Ga_{0.47}As$ guide layers. A first-order grating period was $\Lambda = 1.04$ μm for the single mode DFB emission. The coupling coefficient $\kappa$ was estimated to be ~7 cm$^{-1}$. The wafer was processed into 12-μm-wide ridge structures and buried with a semi-insulating Fe doped InP layer. Subsequently, the upper cladding layer was grown with a 5 μm thick n-InP (Si, $1.5 \times 10^{16}$ cm$^{-3}$) and then followed by a 15 nm thick n$^+$- InP (Si, ~$10^{19}$ cm$^{-3}$) cap contact layer. Finally, the top contacts (Ti/Au) was evaporated and followed by electroplating of a thick 5 μm Au layer on top of the laser structure.

The laser bars were equipped in an epitaxial side-up mounting configuration on a copper block and then cooled in open-loop liquid nitrogen cryostats for the initial device characterization, which was performed in the Hamamatsu laboratories in Japan. Figure 1(b) and (c) show the measured emission results for a THz DFG-QCL with 3-mm-long, 12-μm-wide buried-heterostructure waveguide, operated in CW mode at 78 K. Figure 1(b) shows the mid-infrared emission spectra, in which the rapid-scan measurements (a spectral resolution of 0.2 cm$^{-1}$) were performed for the two mid-infrared pumps as well as the generated THz emission from the DFG-QCL. In this device, we adopted the DFB/FP pumping for generating broadband THz emission via nonlinear frequency mixing between a single mode due to the DFB grating and broadband multi-modes due to the FP cavity, as shown in Fig. 1(b). The position of DFB emission was considerably detuned (~90 cm$^{-1}$) from the peak gain; it is important not to suppress the broadband emission due to FP cavity. Consequently, wide bandwidth of the FP modes and high mid-infrared output power are expected to generate broadband THz frequency. After the DFB laser operation at $\lambda_{DFB}$=6.5 µm, the FP lasing takes place at around $\lambda_{FP}$~6.9 µm. The broad FP spectra were confirmed at the pump current above 500 mA and it could be attributed to broadband gain spectrum in the DAU structure. Figure 1(c) shows the THz emission spectra of the DFB device at different currents in linear scale; the ultra-broadband THz emission with many longitudinal modes ranges from 1.8 THz to 3.3 THz at 78 K, which is a consequence of the frequency down conversion of mid-infrared, multi-mode emission spectra due to the FP cavity.

Current-voltage and light-current characteristics of both mid-infrared and THz power outputs are depicted in Fig. 2. The DFG-QCL demonstrates a mid-infrared CW power of 1.1 W as well as a CW THz power of over 8 µW at 78 K, exhibiting a broadband THz emission (Fig. 1(c)). Figure 2 also shows the temperature dependence of THz CW light-current curves in the different temperatures of 78–170 K. A THz power output of approximately 0.5 µW at 170 K can be noticed. The maximum operating temperature of the present device is significantly lower than our previous papers [40,45], in which room temperature CW operation has been achieved using the epitaxial side-up mounting configuration [45]. This indicates that thermal management of the active region of DFG-QCLs is very difficult for the epitaxial side-up mounted THz DFG-QCLs and an epitaxial side-down mounting technique [39,46,49] is imperative for stable CW operation at room temperature.

## THz Multi-heterodyne detection

A more accurate characterization of the spectral emission of the DFG-QCL has been performed at the CNR-INO laboratories in Italy, by mounting it on the cold finger of a liquid Helium cryostat. In order to perform the measurements at the same device operational temperature as the ones reported in the previous paragraph, we measured the resistance of the DFG-QCL device, which is temperature dependent. In fact, due to the different cryostats configurations, the temperature sensors distance from the device can be very different, leading to inconsistent temperature measurements. In the experiments, the QCL device has been driven by means of an ultra low-noise current driver (ppqSense, QubeCL-P05) at 580 mA. Following the current-voltage curve of the QCL device in Fig. 2, this corresponds, at 78 K, to a resistance of about 23.3 Ω. Therefore, we changed the operational temperature of the device to obtain, at I=580 mA, an applied voltage of 13.5 V, and therefore the same QCL resistance of 23.3 Ω.

On the QCL mount, a bias-tee (Marki Microwave, BT-0024SMG) is connected so as to retrieve the radio-frequency (RF) beatings among the different modes emitted, named hereafter intermodal beatnotes (IBNs). These radio-frequency IBNs are acquired by a spectrum analyzer (Rohde-Schwarz, FSW 26.5 GHz). A frequency-comb operation of the FP mid-infrared QCL device requires the presence of a single IBN frequency ($f_{IBN}$). For the selected driving current ($I_{QCL}$=580 mA) and device resistance ($R_{QCL}$=23.3 Ω), we notice the presence of a single IBN, as reported in Fig. 3 ($f_{IBN}$ = 15.18 GHz) with different spans and resolution bandwidths. Yet, the information retrieved by the intermodal beatnote is not sufficient to confirm a comb-like emission from the FP mid-infrared laser QCL, and, as a consequence, whether the DFG device is emitting a frequency comb in the THz range. To these purposes, we implemented a multi-heterodyne detection setup, detecting the THz beating of the QCL-comb with a well-known reference frequency comb emitting in the THz range.

The THz spectral emission of the DFG-QCL comb is characterized in continous-wave (CW) mode according to a standard multi-heterodyne detection procedure described in [30,31], with the experimental setup sketched in Figure 4. The reference comb is obtained through optical-rectification (OR) of an amplified mode-locked Erbium-doped fiber fs-laser (Menlo Systems, model FC1500) locked against a primary frequency standard, whose emission is focused in a single-mode Lithium Niobate waveguide. The resulting THz frequency comb, hereafter referred as OR-comb, presents several advantages: it is very stable (6 Hz stability), it is offset-free, and it has a repetition rate $f_{rep}$ continuously tunable from 248 to 252 MHz. This OR-comb radiation is mixed with the QCL emission on a Hot Electron Bolometer (HEB-Scontel RS0.3-3T1), realizing multi-heterodyne detection, and retrieving the down-converted beating RF-signal between the OR- and the QCL-combs, consisting of the beating of each optical mode effectively emitted by the QCL with each OR-comb mode.

These heterodyne beatnotes (HBNs) are particularly useful when the ratio between the mode spacings of the two parenting combs are close to an integer value, as illustrated in Figure 5. The HBNs are acquired on a spectrum analyzer (Tektronix, RSA5106A), and two sample spectra are presented in Figure 2. In particular, Figure 6a shows an acquisition with $f_{rep}$ chosen as exact submultiple of $f_{IBN}$. In this configuration, assuming the DFG-QCL behaves as a comb, the frequency differences between each QCL-comb mode and its closest neighboring OR-comb mode are exactly the same. As a consequence, in the down-converted RF spectrum, all the HBN have to collapse at the same frequency, which is exactly the case of Fig. 6a, confirming the comb-like nature of the DFG device. In Figure 6b, $f_{rep}$ is slightly detuned, and we can visualize 5 HBNs corresponding to 5 THz modes, equally spaced in the frequency domain, with a 40 MHz span and a 10 Hz resolution bandwidth (RBW). These HBNs correspond to the most intense THz modes emitted by the QCL device around 2.4 THz, as shown in Fig. 1b. The HBNs signal levels are between 5 and 10 dBm, and these signal to noise ratios do not allow a characterization of the level of coherence of the emitted comb, i.e. application of the Fourier analysis of comb emission (FACE) technique. Yet, the realized experimental setup and the retrieval of the HBNs allow to measure frequencies of all the QCL modes with a very high precision. In fact, if we take into account the frequency of the most intense QCL mode N ($f_N$), corresponding to the most intense HBN shown in Fig. 6, we can write down the equation

$$f_N = M \cdot f_{rep} \pm f_{HBN} \qquad \text{(eq. 1)}$$

where $f_{HBN}$ is the frequency of the HBN signal between the $N^{th}$ QCL mode and the $M^{th}$ OR-comb mode.

By fixing constant values for $I_{QCL}$ and $R_{QCL}$, the QCL mode frequency $f_M$ remains constant. Then, by modifying the OR-comb repetition rate $f_{rep}$ and tracking the $f_{HBN}$ frequency, from a simple linear regression we can extrapolate the order M of the OR-comb mode, as shown in Figure 7. The linear regression of this dataset (green line) results in a precise estimation of the order M. In fact, since the retrieved value is 9705.01 with a 0.33 standard deviation, and since M is integer, we can round to 9705 as mode number M of the OR-comb. Therefore, the exact M order is used for the $f_N$ QCL mode frequency determination. Indeed, in equation (1), where the QCL mode frequency is calculated, the only remaining sources of uncertainties are on the values of $f_{rep}$ and $f_{HBN}$. The latter, with the 1.0 MHz linewidth observed in Figure 6b, being predominant. As a consequence, the QCL frequency is determined as 2416068.1(1.0) MHz. We can then use the IBN value and the order of the various modes (as seen in Fig. 6) to simultaneously measure the frequencies of all the emitted modes, confirming that these modes correspond to the most intense shown in Fig. 1b, acquired with the FTIR spectrometer. With respect to those measurements, the multi-heterodyne technique permits to increase the accuracy in the retrieval of modes' frequencies by more than 3 orders of magnitude and, more importantly, allows a simultaneous measurement for all the detected QCL comb modes. As a matter of fact, it can be noticed that by performing faster acquisitions on a smaller frequency window, the accuracy on the frequency of a single HBN can be further improved down to the kHz level, but this would compromise the simultaneous acquisition of all the emitted modes.

## Conclusions

In conclusion we have presented the first direct observation of the THz modes emitted by a DFG-QCL frequency comb by multiheterodyne detection with an optically rectified THz comb, referenced to the primary frequency standard, that is a 10-MHz quartz-oscillator disciplined by a Rb-GPS (Global Positioning System) clock (stability of $6\times10^{-13}$ in 1 s and absolute accuracy of $2\times10^{-12}$). Thanks to our setup, the THz QCL comb modes frequencies could be simultaneously retrieved with a 1 MHz accuracy, confirming that they are actually equally spaced in frequency, at this level of precision. Moreover, by tuning the ratio between the mode spacings of the two combs to an integer value, we observe the HBN collapsing, as expected from a frequency comb.

Unfortunately, due to the low emitted power-per-mode, only 7% of the total modes could be down-converted, and state-of-the-art techniques such as FACE could not be applied to the THz emission. This could have confirmed the existence of a fixed phase relation among the modes, that is ultimately the true frequency comb signature. However, further improvements in output power can be achieved by the adoption of an optimized device structure with a long-wavelength nonlinear active region [42] and silicon-based THz waveguide [52]. Once the power issues are overcome, not only the FACE technique can confirm phase coherence in the THz emitted modes, but also this technology can fully deploy its promising potential for broadband precision spectroscopy in the THz range, even without cryogenic cooling systems. The latter would pave the way for miniaturized precise metrology diagnostics in the THz range.

## Acknowledgements

This work was supported by MIC/SCOPE #195006001, Qombs Project, EU-H2020-FET Flagship on Quantum Technologies grant no. 820419, the Italian ESFRI Roadmap (Extreme Light Infrastructure-ELI), EC–H2020 Laserlab-Europe grant agreement 871124. Authors at Hamamatsu Photonics thank Naota Akikusa for his helpful support and comments, KF expresses his thanks to Prof. M. A. Belkin of the Technical University of Munich, and Dr. S. Jung of TransWave Photonics, LLC for helpful discussions.

## Data availability

The data that support the findings of this study are available from the corresponding author upon reasonable request.

**Figures:**

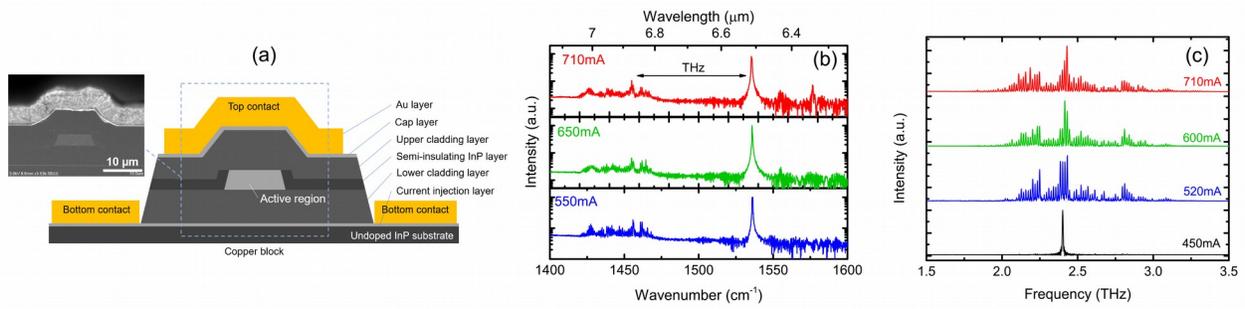

Figure 1: (a) Schematic of the THz DFG-QCL and Scanning electron microscope (SEM) image of the device. Mid-infrared (b) and THz (c) spectra at different currents of the DFG-QCL at the temperature of 78 K used in this study. The cavity length is 3 mm and the ridge width is 12 μm.

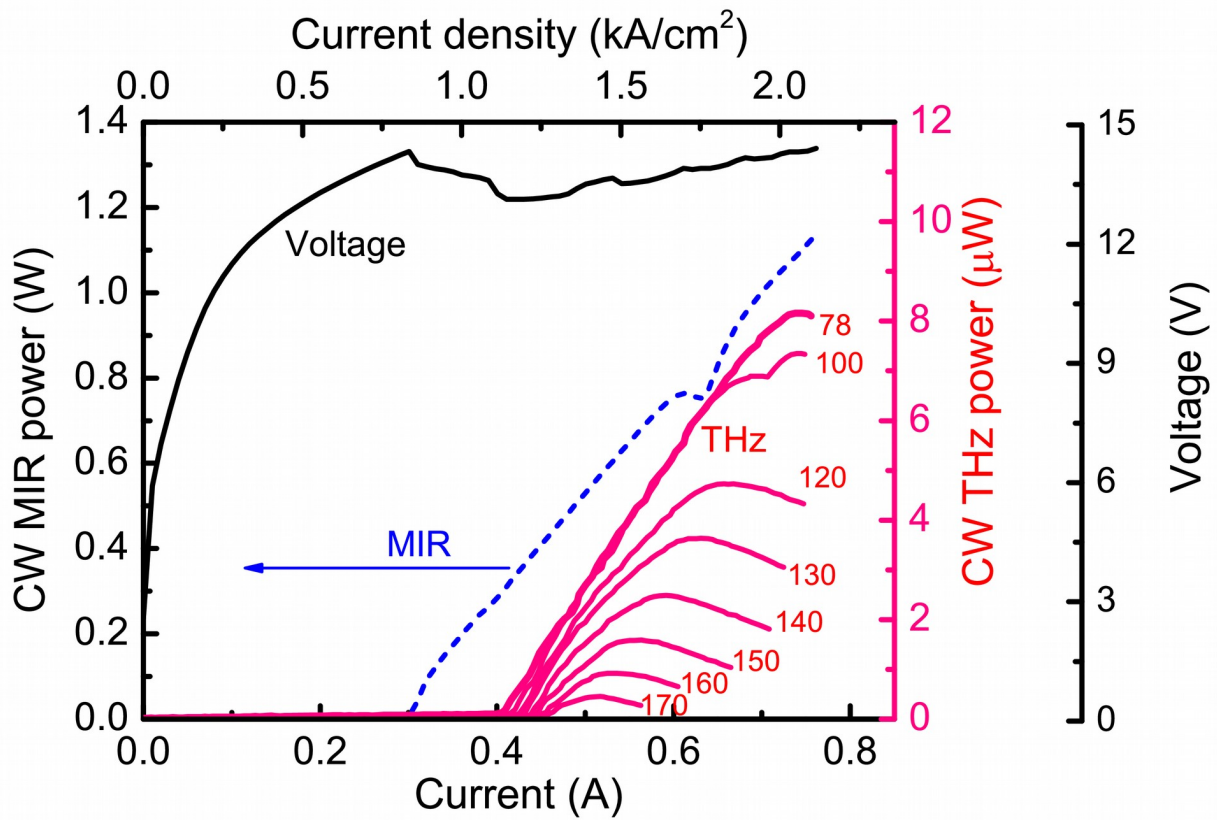

Figure 2: Current-voltage and light-current characteristics of the mid-infrared pumps at 78 K and THz DFG at various temperatures, for the device operated in CW mode.

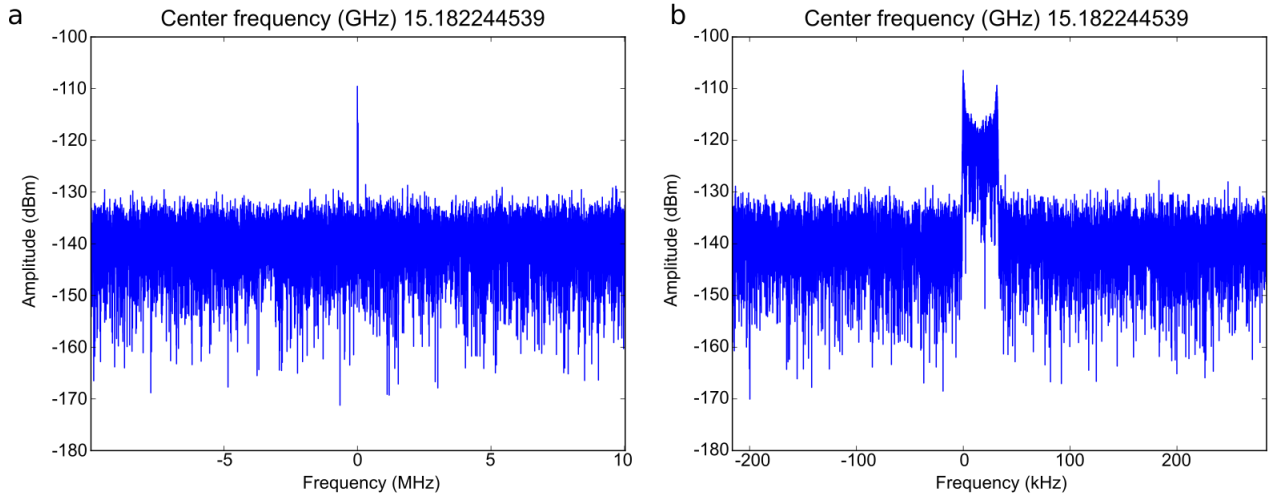

Figure 3: Observation on the spectrum analyzer of a single internodal beatnote (IBN) for a 580 mA driving current and 23.3 Ω device resistance: (a) Span = 20 MHz, RBW = 100 Hz ; (b) Span = 20 MHz, RBW = 50 Hz. RBW : resolution bandwidth.

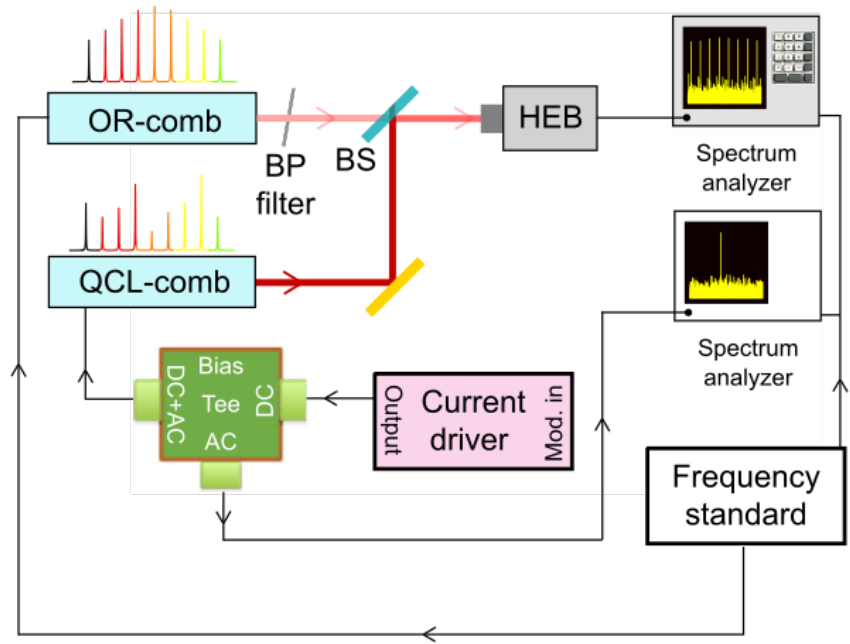

Figure 4: Experimental setup used for the characterization of the DFG QCL-comb. The beams of the optically-rectified comb (OR-comb) and QCL-comb are superimposed by means of a beam splitter (BS) and then mixed on a fast detector (HEB: hot-electron bolometer). The HEB signal is acquired on a spectrum analyzer (Tektronics RSA5106A), and the intermodal beatnote (IBN) is acquired on a second spectrum analyzer (Rohde Schwarz FSW 26.5 GHz). The OR-comb and both spectrum analyzers are frequency-referenced to the primary frequency standard. BP filter: band-pass filter

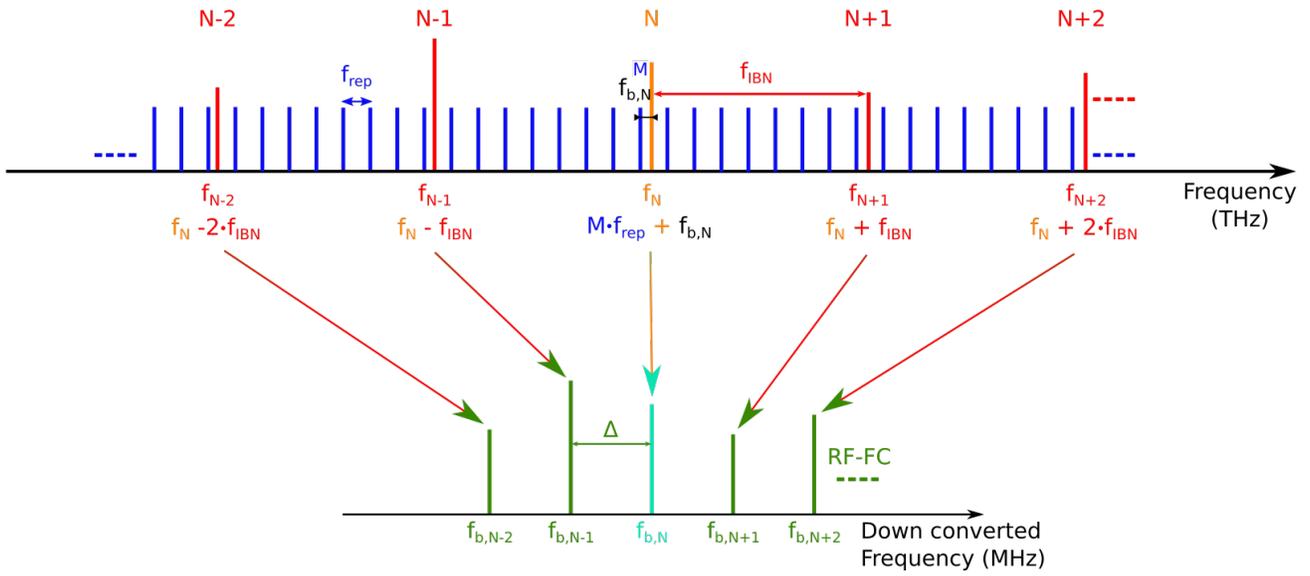

Figure 5: Illustration of the multi-heterodyne down-conversion process. Schematic representation of the quantum cascade laser (QCL-) (red) and optically rectified (OR-) frequency comb (FC) (blue), whose modes are respectively spaced by $f_{IBN}$ and $f_{rep}$. These two repetition frequencies are tuned close to an integer ratio, allowing an ordered and distinguishable down-conversion to radio frequencies (RF). In fact, the down-converted RF-FC modes (green) are equally spaced by $\Delta$, and their easily measurable RF frequencies are used to calibrate the absolute frequency scale of the QCL-FC, as described in the main text. Reprint with permission from [36] Copyright Communications Physics.

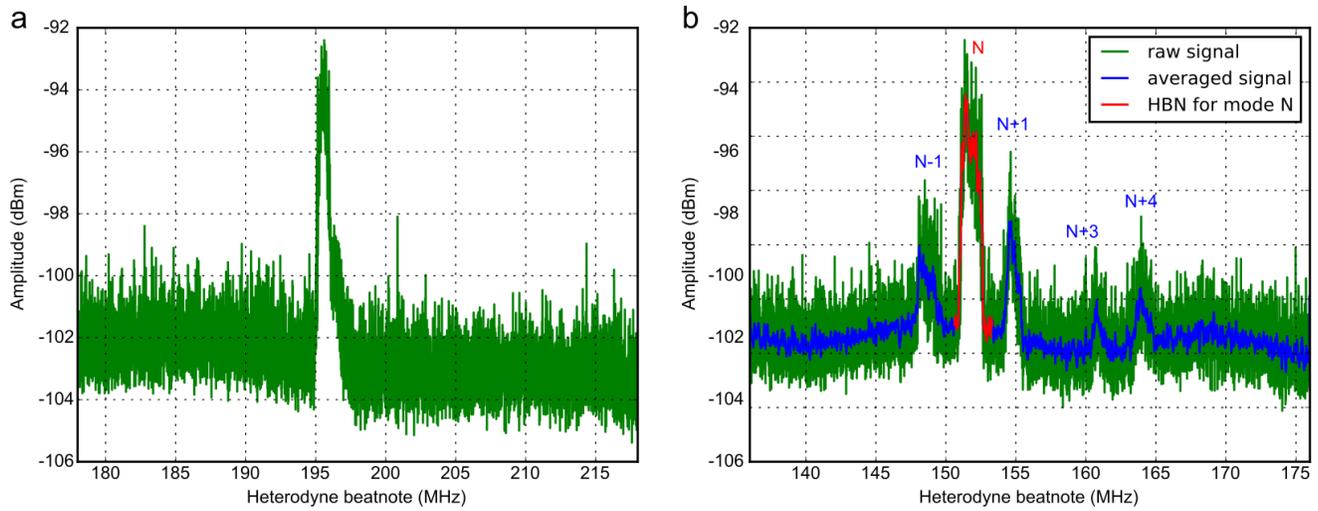

Figure 6: Acquisition of the heterodyne beatnotes signal (HBN) on the spectrum analyzer, resulting from the mixing of the OR- and QCL- frequency combs, characterized by their intermodal frequencies $f_{rep}$ and $f_{IBN}$, on the Hot-Electron Bolometer. The HBNs are: (a) collapsed when $f_{rep} = (f_{rep})_0$, submultiple of $f_{IBN}$ ; (b) corresponding to the optical modes when $f_{rep}$ is slightly detuned from $(f_{rep})_0$, where modes are equally spaced by $\Delta$. Resolution bandwidths = 10 Hz, blue: averaged signal, red: most intense mode M highlighted.

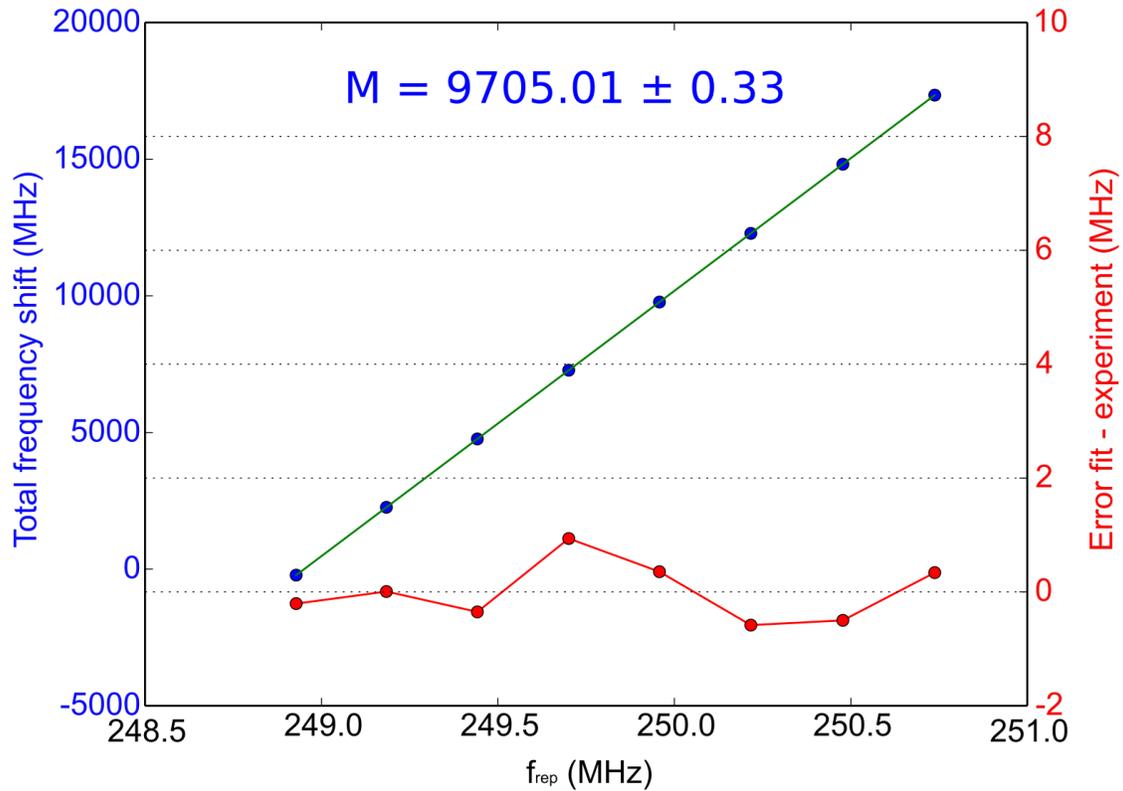

Figure 7: Retrieval of the optically rectified frequency comb (OR-comb) mode number M relative to eq. 1. The frequency of the QCL-comb mode involved in the beating is kept constant by fixing the device driving current and operational temperature. The frequency of the down converted beatnote HBN is acquired and plotted as a function of the optically rectified comb repetition rate ($f_{rep}$). As a consequence, the order M is extracted from the slope of the data linear regression, by rounding to the nearest integer, i.e., M = 9705. The fit residuals, plotted in red, confirm the 1 MHz uncertainty in the determination of the $f_{HBN}$ frequency.